\documentclass[fleqn,usenatbib]{mnras}

\usepackage{newtxtext,newtxmath}
\usepackage[T1]{fontenc}
\usepackage{ae,aecompl}

\usepackage{graphicx}	% Including figure files
\usepackage{amsmath}	% Advanced maths commands
\usepackage{amssymb}	% Extra maths symbols

\usepackage{subfig}

\usepackage{booktabs}

\usepackage{cleveref}
\crefname{section}{\S}{\S\S}
\Crefname{section}{\S}{\S\S}

\newcommand{\Lya}{Ly-${\alpha}$}
\newcommand{\Lyb}{Ly-${\beta}$}
\newcommand{\Lyg}{Ly-${\gamma}$}

   \title[Discrete galaxies at cosmic dawn]{Mapping discrete galaxies at cosmic dawn with 21-centimeter observations}
\author[Reis et al.]{
Itamar Reis$^{1}$\thanks{E-mail: itamarreis@mail.tau.ac.il},
Rennan Barkana$^{1}$,
and Anastasia Fialkov$^{2}$
\\
$^{1}$School of Physics and Astronomy, Tel-Aviv University, Tel-Aviv, 69978, Israel\\
$^{2}$Institute of Astronomy, University of Cambridge, Madingley Road, Cambridge CB3 0HA, UK\\
}

\date{Accepted XXX. Received YYY; in original form ZZZ}

\pubyear{2020}

\begin{document}
\label{firstpage}
\pagerange{\pageref{firstpage}--\pageref{lastpage}}
\maketitle

\begin{abstract}
At cosmic dawn, the 21-centimeter signal from intergalactic hydrogen was driven by \mbox{Lyman-$\alpha$} photons from some of the earliest stars, producing a spatial pattern that reflected the distribution of galaxies at that time. Due to the large foreground, it is thought that around redshift 20 it is only observationally feasible to detect 21-cm fluctuations statistically, yielding a limited, indirect probe of early galaxies. Here we show that 21-cm images at cosmic dawn should actually be dominated by large (tens of comoving megaparsecs), high contrast bubbles surrounding individual galaxies. We demonstrate this using a substantially upgraded semi-numerical simulation code that realistically captures the formation and 21-cm effects of the small galaxies expected during this era. Small number statistics associated with the rarity of early galaxies, combined with the multiple scattering of photons in the blue wing of the \Lya{} line, create the large bubbles and also enhance the 21-cm power spectrum by a factor of 2--7 and add to it a feature that measures the typical brightness of galaxies. These various signatures of discrete early galaxies are potentially detectable with planned experiments such as the Square Kilometer Array or the Hydrogen Epoch of Reionization Array, even if the early stars formed in dark matter halos with masses as low as $10^8\, M_\odot$, ten thousand times smaller than the Milky Way halo.
\end{abstract}


%%%%%%%%%%%%%%%%%%%%%%%%%%%%%%%%%%%%%%%%%%%%%%%%%%

%%%%%%%%%%%%%%%%% BODY OF PAPER %%%%%%%%%%%%%%%%%%

\section{Introduction}
The beginning of cosmic dawn is the most exciting target of 21-cm observations since it is the earliest period with a strong signal that is feasible to probe with upcoming 21-cm experiments \citep{madau97}. During this period, high-redshift galaxies drove the 21-cm signal as the non-ionizing ultraviolet photons emitted from stars were redshifted by cosmic expansion to the nearest Lyman-band frequency. Many photons reached the \Lya{} frequency (directly or by cascade), near which the photons were absorbed and re-emitted hundreds of thousands of times by intergalactic atomic hydrogen before being redshifted out of the line \citep{barkana16,Mesingerbook}. During this absorption and re-emission, through the subtle Wouthuysen - Field effect \citep{wouthuysen52, field58} these photons drove the spin temperature (defined as the effective temperature describing the occupation ratio of hyperfine levels in the ground state of hydrogen) very close to the kinetic temperature of the gas. This is in contrast with the earlier approximate equality between the spin temperature and the cosmic microwave background (CMB) temperature, an equality that was broken as a result of the formation of the first stars. This so-called \Lya{} coupling transition is expected to be observable as a prominent 21-cm absorption feature since the kinetic temperature was much lower than the CMB temperature at these redshifts. The two essential ingredients in simulating the fluctuations of the 21-cm signal during the coupling transition are the clustering properties of galaxies, which were the sources of the \Lya{} photons, and the distribution around these sources of the photons that were absorbed and thus produced the coupling effect. X-ray heating and other astrophysical effects were likely insignificant at this early time (see below).
 
Accurate predictions of the 21-cm signal at high redshift require us to follow the evolution of large volumes ($> 100$~Mpc on a side), for several reasons: the radiation (including \Lya) that drove the 21-cm signal reached out to these distances from each source; upcoming observations will be limited by resolution and other constraints to imaging scales from $\sim 10$~Mpc to a few hundred; and most importantly, the first galaxies represented rare peaks in the cosmic density field, leading to surprisingly large fluctuations in their number density on large scales \citep{barkana04}, which drove observable 21-cm fluctuations during the \Lya{} coupling era \citep{barkana05}. Full numerical simulations that capture these large scales cannot resolve the small halos expected to dominate star formation at cosmic dawn. Indeed, large-scale simulations run at cosmic dawn \citep{Ghara17,semelin17} can resolve halos down to a mass of $4 \times 10^9~M_\odot$ at best. Instead, most simulations focus on the later era of cosmic reionization \citep{CODA} or on achieving sufficient resolution within much smaller simulated volumes \citep{ahn15,xu16} \citep[generally too small for 21-cm cosmology, ][and more specifically, smaller than a {\it single} bubble of the type we present below]{barkana04}. At the other extreme, fully analytical approaches have often been used to introduce novel ideas including \Lya{} fluctuations \citep{barkana05}, but such calculations require crude approximations (usually including the assumption of small, linear fluctuations in many quantities), and so are too inaccurate. Thus, the most realistic predictions of the 21-cm signal from cosmic dawn have come from various intermediate methods termed semi-numerical simulations \citep{mesinger11,visbal12,kaurov18,munoz19}, which combine analytical models normalized to the results of full simulations on small scales, with a detailed numerical integration of the relevant radiation fields on large scales. 

Semi-numerical simulations usually generate galaxy distributions based on models \citep{press74, sheth99, barkana04} for the mean number of dark matter halos per volume, given the density field. This approach produces a good fit to the average number of halos found in full numerical simulations, but it fails to capture the Poisson fluctuations (shot noise), which are necessarily prominent at sufficiently high redshifts, when the number density of star-forming halos was low. The possible role of Poisson fluctuations in 21-cm observations at cosmic dawn has been previously investigated with approximate analytical calculations \citep{barkana05} or with approximate simulation-based methods \citep{kaurov18} that resolved halos of mass $2 \times 10^9\ M_\odot$ and higher. Sub-grid methods have been used to approximately insert low-mass halos into full radiative-transfer simulations that investigated cosmic heating and reionization \citep{ross17}, but not the \Lya{} era (the saturated coupling limit was assumed). Publications that use the semi-numerical code \texttt{21cmFast} \citep{mesinger11} apparently do not include Poisson fluctuations. The \texttt{SimFast21} code has been run with Poisson-generated halos but only for a case dominated by low-mass halos \citep{santos11}, in which the signatures we highlight here are too weak to be observable. 

We have modified an existing semi-numerical simulation \citep{visbal12, fialkov14a, cohen16} to fully incorporate the shot-noise contribution to the clustering of galaxies, for all halo masses including those expected to dominate star formation at cosmic dawn. To do this we generated all star-forming halos individually from a Poisson model centered on the expected mean distribution of halo masses. It is important to keep an open mind and consider a wide range of possible galactic halo masses, as we do below, since on the one hand, low-mass halos are the most abundant at early times, but on the other hand, efficient star formation may occur only in massive halos, as suggested by both extrapolations of low-redshift observations \citep{mirocha} and the results of numerical simulations that achieve high resolution in small volumes \citep{xu16}.
 
The second ingredient necessary for realistic predictions of the 21-cm signal from the \Lya{} coupling era is the radiative transfer of the photons. The path that \Lya{} photons travel through the intergalactic hydrogen, from emission through cosmological redshift and until absorption near the \Lya{} line center, is commonly approximated by a straight line. In reality, photons emitted in the range of frequencies between \Lya{} and \Lyb{} usually scatter from the blue wing of the \Lya{} line, long before reaching the line center. Such multiple scattering results in photons traveling shorter effective distances from their sources before absorption, compared to the no-scattering approximation. While the \Lya{} photons can travel up to hundreds of Mpc, multiple scattering creates an over-concentrated halo of \Lya{} photons at a characteristic comoving distance (from where the line center is reached) of \citep{loeb99} 
\begin{equation}
    R_* = 21 \times \left(\frac{\Omega_b / \Omega_m}{0.157}\right) \left(\frac{1+z}{20}\right)\  {\rm Mpc}\ ,
\end{equation}at redshift $z$. Analytical and numerical calculations \citep{chuzhoy07b,semelin07,naoz08,vonlanthen11,higgins12} have suggested that this should boost 21-cm fluctuations, but large-scale simulations \citep{vonlanthen11,semelin17} that incorporate radiative transfer of \Lya{} are severely limited, only resolving halos above a mass of $9 \times 10^{10}\ M_\odot$. In order to make realistic predictions for the halo masses expected to host galaxies at cosmic dawn, we have added this effect to our semi-numerical simulations, using a Monte-Carlo calculation of the effective distance distribution of \Lya{} photons as a function of the emission and absorption redshifts. 

This paper is organized as follows: In \cref{sec:methods} we present our semi-numerical 21-cm simulation and discuss the modifications introduced in this work. In \cref{sec:results} we show the results of the upgraded simulation, focusing on cosmic dawn. We summarize in \cref{sec:summary}.

\section{Methods}
\label{sec:methods}

In this work we have extended an independent 21-cm semi-numerical simulation code that we previously developed \citep{visbal12, fialkov14a, cohen17}. The approach used in our code  was originally inspired by \texttt{21cmFast} \citep{mesinger11}. Our code simulates the evolution of the 21-cm signal in a 3-dimensional volume composed of 128 voxels on a side, each with a  size of 3 comoving Mpc. The simulation produces a realization of the 21-cm signal from cosmic dawn, arising from the coupling transition due to \Lya{} photons from the first stars ($z \sim 20-30$), through the heating of the intergalactic medium by the first X-ray sources, until cosmic reionization ($z \sim 6-10$). In this work we have focused on the high-redshift coupling transition, the earliest era of galaxy formation that is feasible to detect with upcoming observations. While our simulation includes  heating of the IGM by X-ray photons, and reionization by UV photons, these do not play an important role in the models we consider here.

\subsection{Simulating the high-redshift galaxy population}

The first step of the simulation is obtaining a sample  of dark matter halos in the simulation volume. We start by creating a random realization of the large-scale, linear, density field, given its statistical properties (specifically, the power spectrum of the initial Gaussian random density field) as measured by the Planck satellite \citep{planckcollaboration18}. Note that fluctuations on the scale of the voxel size (3 comoving Mpc) are still rather linear at the high redshifts considered. Given the large-scale, linear density field, we obtain the population of collapsed dark matter halos inside each voxel, using a modified Press-Schechter model \citep{press74, sheth99, barkana04} that was fitted to match the results of full cosmological simulations. 

A major modification to this procedure, introduced in this work, is adding Poisson fluctuations to the number of halos predicted by the modified Press-Schechter model. In each time-step of the simulation, we calculate the predicted number of {\it new halos formed in the time step}, in different mass bins, and draw the created halos from a Poisson distribution with the predicted number acting as the mean. Adding Poisson fluctuations to the number of halos created in each time step allows us to create a complete realization of the time evolution of the 21-cm signal. While this simplified calculation of the halo population neglects correlations between different mass bins, it is sufficient for the era we focus on here, where almost every pixel in the box has either a single galactic halo or none. As noted in the introduction, publications that use the semi-numerical code \texttt{21cmFast} do not seem to include Poisson fluctuations; e.g., Fig.~3 of a paper \citep{Kern17} that used \texttt{21cmFast} shows an example that corresponds to $V_c \sim 37$~km/s and $f_* = 0.1$, yet at the high-redshift end the 21-cm power spectrum is flat and does not show the break that we find due to Poisson-enhanced individual halo bubbles. More clearly, in a recent paper \citep{park19} that added into \texttt{21cmFast} a duty cycle for star-forming halos, the duty cycle was inserted as a simple multiplicative factor into the various radiative emissivities, with no mention of the additional effect of an increase in Poisson fluctuations that would be found in a code that {\it did}\/ include individual halos.

Given a dark matter halo of mass $M$, the baryon fraction contained in the halo is assumed to be the cosmic mean, except for a reduction due the streaming velocity \citep{tseliakhovich11, fialkov12}. A halo forms stars if $M > M_{\rm min}$ where $M_{\rm min}$ is the minimum mass for star formation, determined by gas cooling and/or feedback.  In this paper this minimum mass for star formation is parameterized by the circular velocity (a more direct measure of the depth of the potential well, and also the virial temperature), defined as the velocity of a circular orbit at the halo virial radius. For a halo of mass $M$,
\begin{equation}
    V_c = 16.9 \left(\frac{M}{10^8 M_{\odot}}\right)^{1/3}\left(\frac{1+z}{10}\right)^{1/2}  \left(\frac{\Omega_m h^2}{0.141}\right)^{1/6} \left(\frac{\Delta_c}{18 \pi^2}\right)^{1/6}\   {\rm km}\;{\rm s}^{-1}\ ,
\end{equation}
where $\Delta_c$ is the ratio between the collapsed  density and the critical density at the time of collapse, which equals $18 \pi^2$ for spherical collapse.

The stellar mass M$_{\star}$ in each star-forming halo is the gas mass times the star formation efficiency f$_{\star}$. The luminosity of the galaxy is assumed to be proportional to the star formation rate (SFR). We apply two commonly-used approaches to obtain the SFR from M$_{\star}$ \citep{mesinger11,park19}: 
\begin{equation}
    {\rm SFR} = \frac{d{\rm M}_{\star}}{dt}\ ,
\end{equation}
corresponding to a bursting mode in newly-accreted gas, and
\begin{equation}
    {\rm SFR} = \frac{{\rm M}_{\star}}{t_{\star} H(z)^{-1}}\ ,
\end{equation}
corresponding to a quiescent mode in previously-accreted gas. Here $H(z)^{-1}$ is the Hubble time, and $t_{\star}$ is an additional parameter that we set to $0.2$ so that $t_{\star} H(z)^{-1}$ corresponds approximately to the characteristic dynamical time of a halo (at its virial density). We have performed tests using each of the two star-formation modes separately, and found that while these two SFR prescriptions result in a somewhat different time evolution of the SFR, there is no significant difference to the coupled bubbles picture. Since in reality both modes are likely to contribute, in our examples here we have assumed that the total SFR is given by the sum of the two modes (and then the bursting mode usually dominates at the redshifts considered here). 

In the examples shown in this work we have assumed $f_* = 0.1$ as our standard value, and used $f_* = 0.3$ to illustrate a case with higher $f_*$. The increased abundance of star-forming halos in models with very low $V_c$ allows such models to reach 21-cm milestones at reasonable redshifts with lower $f_*$, so we illustrated these models with $f_* = 0.03$ for $V_c = 16.5$~km/s, and $f_* = 0.007$ for $V_c = 4.2$~km/s. For $V_c$ we have used characteristic values for molecular hydrogen cooling ($4.2$~km/s), atomic cooling ($16.5$~km/s), ten times the halo mass of atomic cooling ($35.5$~km/s), 100 times the atomic cooling mass ($76.5$~km/s), and additional intermediate values ($25$ and $50$~km/s). For the excess-radio model \citep{fialkov19} we set $f_* = 0.1$ and the radio background assumed a Galactic-like synchrotron spectrum that has an amplitude three times the CMB brightness temperature at 78~MHz. 

\subsection{Full numerical simulations of cosmic dawn}

Accurate predictions of the 21-cm signal at high redshift require us to follow the evolution of large volumes for several important reasons. The required volumes begin from a minimum of $100$~Mpc on a side, but a more advisable number is a few hundred Mpc. N-body simulations in which the dark matter halo distribution is realistically generated can achieve a minimum resolved halo of $4 \times 10^9~M_\odot$ in a volume that is 430~Mpc on a side \citep{Ghara17}. These are gravity-only numerical simulations on top of which approximate methods are later used to add star formation and the various astrophysical radiation fields. Approximate simulation-based methods \citep{kaurov18} that were used to explore Poisson fluctuations have been able to generate halos of mass down to $2 \times 10^9\ M_\odot$ in a volume 910~Mpc on a side. These works all assumed optically thin \Lya{} evolution (with no scattering except at the center of the \Lya{} line). Numerical simulations run at cosmic dawn with numerical radiative transfer of the \Lya{} photons \citep{semelin17} resolved halos of mass $8 \times 10^8~M_\odot$ in a volume 30~Mpc on a side (barely able to capture the \Lya{} halo around a single galaxy), or $9 \times 10^{10}~M_\odot$ in a box of side 150~Mpc.

In listing these various numbers of minimum resolved halos, we have adopted the common assumption of 20 simulation particles needed in order to resolve a halo. This number, however, is quite optimistic. Numerical resolution studies \citep{springel03} suggest that $>100$ particles are necessary in order to determine even the overall mass of an individual halo to within a factor of two, and even more particles are needed for quantities such as the overall star formation rate in the halo (which is sensitive to the merger history and thus to the small precursor halos that are less well-resolved). 

\subsection{\Lya{} photon distance distribution}

Given the population of galaxies obtained as described above, we calculated the spatial distribution of the \Lya{} photons that they produce. As explained in the introduction, \Lya{} photons were the driver of the early 21-cm signal at cosmic dawn. These \Lya{} photons originated as continuum photons emitted at frequencies between \Lya{} and the Lyman limit. The emitted photons produced \Lya{} photons by two different mechanisms: (i) Photons emitted at frequencies between \Lya{} and  \Lyb{} were redshifted directly to  the \Lya{} frequency by cosmic expansion,  and (ii) photons emitted at higher frequencies were absorbed in higher Lyman series frequencies and created atomic cascades;  $\sim 30 \%$ of  cascades originating from \Lyg{} and above produced \Lya{} photons, while no \Lya{} photons were produced by cascades originating from \Lyb{} \citep{hirata06, pritchard06}.  In this work, the distribution of emitted photons is calculated assuming Population II stars, while Population III stars would lower the \Lya{} output by about a factor of two \citep{barkana05,bromm01} (while substantially increasing the ionizing photon output, which is unimportant at the redshifts that we consider here); such a change is nearly degenerate with a change in $f_*$ (only nearly because of the effect of the stellar spectrum which, however, is small due to the narrow relevant frequency range).

In previous semi-numerical simulations, the intensity of \Lya{} photons was calculated with the assumption that photons travel in a straight line between emission and absorption at the line center. This assumption made it easy to find the \Lya{} intensity at a point by integrating over previous redshifts, where at each redshift sources contribute only at a single distance from the final arrival point. Thus, the contribution of sources at redshift $z_{\rm emission}$ to the distribution of \Lya{} photons at a lower redshift $z_{\rm absorption}$ was found by convolving the distribution of sources at $z_{\rm emission}$ with a spherical shell window function, with a radius corresponding to the distance that photons travel between $z_{\rm emission}$ and $z_{\rm absorption}$.

\begin{figure*}
    \centering
 {\includegraphics[width=0.49\textwidth]{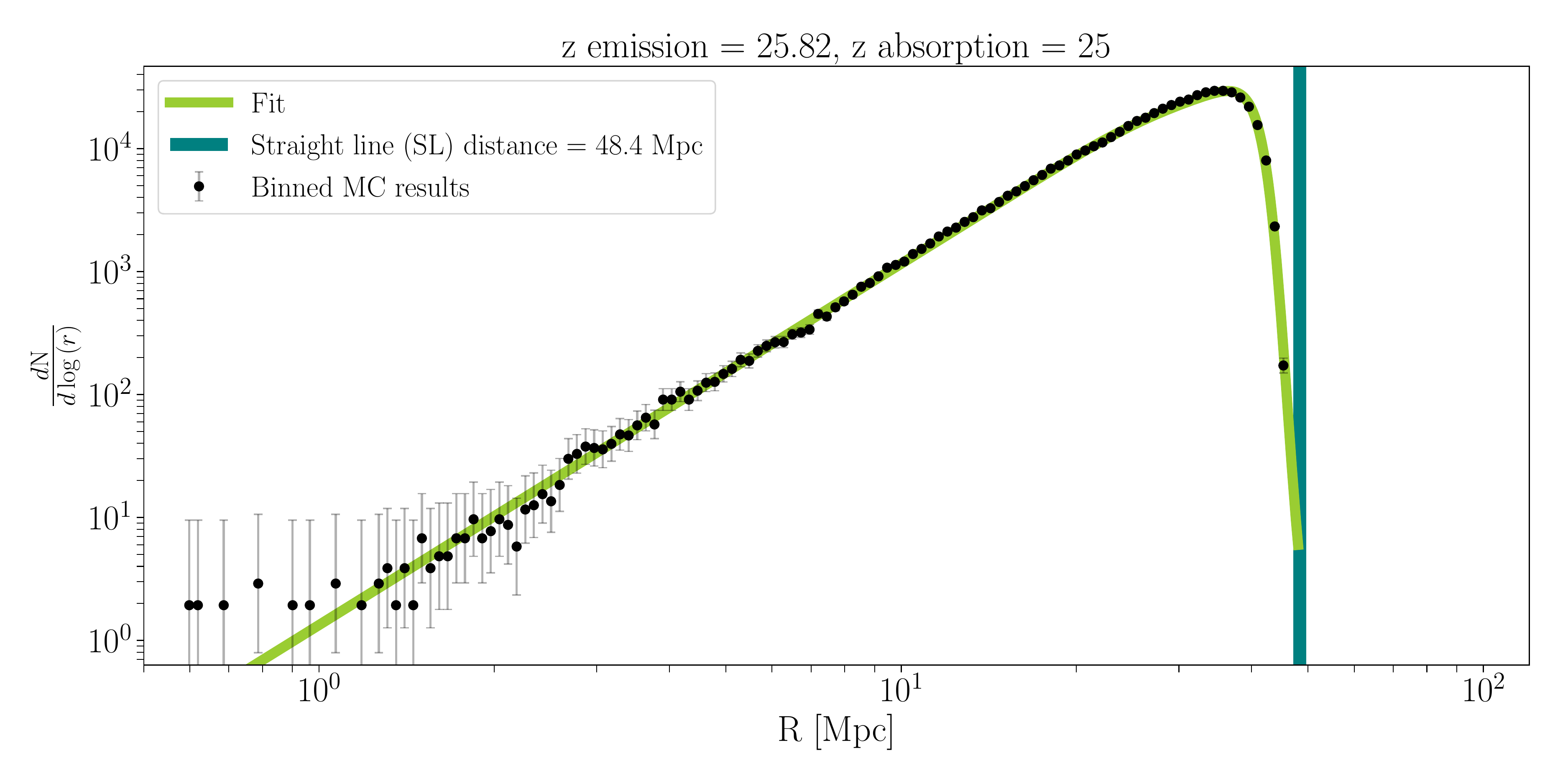}} 
  {\includegraphics[width=0.49\textwidth]{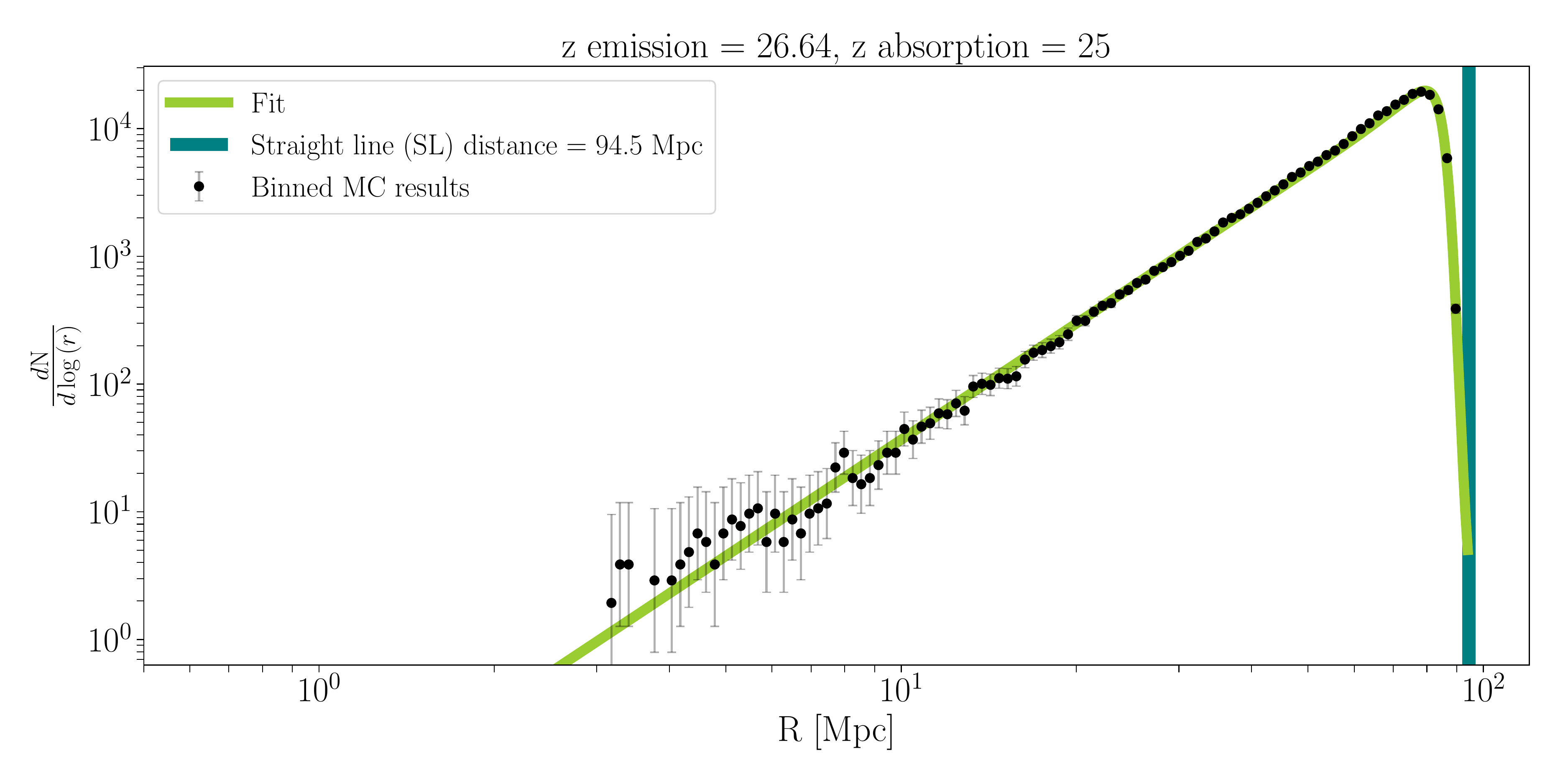}} \\[4pt]
   {\includegraphics[width=0.49\textwidth]{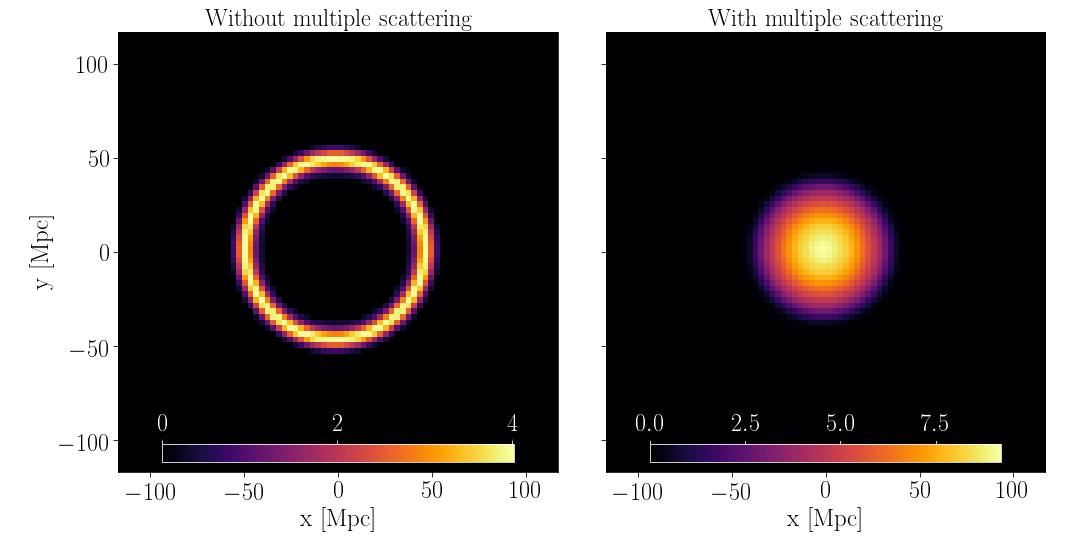}}  
 {\includegraphics[width=0.49\textwidth]{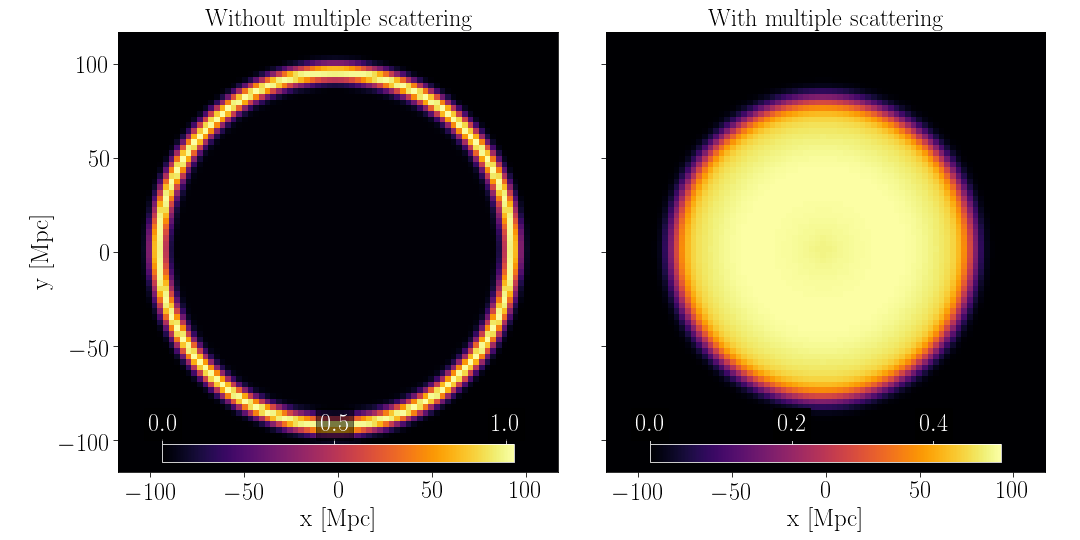}}
    \caption{{\bf Calculation of the multiple scattering of \Lya{} photons.} {\bf Top panels:} Example distributions of the distance from the source at which photons are absorbed at the \Lya{} frequency given an emitted and absorbed redshift, generated using a photon scattering Monte Carlo code. The black dots show the number of photons per bin of log distance, as obtained by the Monte Carlo code, for a total of 250,000 photons per panel. The light-green line shows our fit to the distance distribution, while the dark-green vertical line shows the straight-line distance that all these photons would have traveled without the effect of multiple scattering. {\bf Bottom panels:} The corresponding window functions that are used in our semi-numerical simulation of the 21-cm signal. The window function represents the distribution of photons per volume emitted from a point source at the center (normalized for display purposes to a volume integral of $10^6$). Multiple scattering substantially changes the window functions from the previously-used spherical-shell window functions, which are shown for comparison. All panels show photons emitted and absorbed at specific redshifts: For all panels $z_{\rm absorption} = 25$, with $z_{\rm emission} = 25.82$ for the panels on the left and $z_{\rm emission} = 26.64$ on the right. Fully incorporating these window functions into our code and exploring a wide range of possible astrophysical parameters allows us to go well past previous investigations of the effect of \Lya{} scattering \citep{chuzhoy07b,semelin07,naoz08,vonlanthen11,higgins12}.} 
    \label{fig:example_fit_res}
\end{figure*}

Instead, photons actually scatter elastically off hydrogen atoms in the blue wing of the \Lya{} line before reaching the line center, and thus reach $z_{\rm absorption}$ in a distribution of distances from their source, for any given $z_{\rm emission}$. The straight-line distance is the upper limit of this more realistic distribution that is found when multiple scattering is accounted for. This effect is important for photons emitted between \Lya{} and  \Lyb{}, but not for \Lya{} photons injected from the higher Lyman lines, since the effective distance corresponding to the wing of the line is very small for those transitions \citep{naoz08}. In this work we include multiple scattering by first calculating the effective distance distribution for photons emitted between \Lya{} and  \Lyb{}, using a Monte Carlo code inspired by previous work \citep{loeb99,naoz08}. We then construct a window function that gives a good fit to this distance distribution, and use it instead of the previously used simple, spherical shell window function. We run the Monte Carlo code and construct the window function separately for each combination of emission and absorption redshifts. Two examples are shown in Fig. \ref{fig:example_fit_res}. 

We note that only in our group the semi-numerical simulations since early on  \citep{Complete} have included a rough approximation to the effect of multiple scattering on the 21-cm power spectrum based on an analytical study \citep{naoz08}; while this did boost the power spectrum we have now found that it underestimated the boost by a typical factor of 1.5 and did not capture the correct dependence on wave number or on the astrophysical parameters. We also note that while X-ray heating and UV ionizing radiation do not play a significant role in the 21-cm signal at the early times that we have focused on, these effects are included in our semi-numerical simulations. \Lya{} photons themselves can contribute to the heating of the IGM, but this effect is small during the cosmic dawn, as shown previously  \citep{furlanetto06b} and as we have also verified with our simulation (but note that \Lya{} heating can become important at later redshifts, when the \Lya{} intensity field is significantly larger than required to produce the WF effect). 

\section{Results}
\label{sec:results}

In this work we present the combined effect of Poisson fluctuations and multiple scattering of \Lya{} photons on the 21-cm signal, over a wide range of astrophysical parameters that have never been probed this realistically before. Including these effects in our simulation, we obtain a cosmic dawn 21-cm signal that is substantially different from previous predictions without these effects (Fig.~\ref{fig:vc50_slices}; see Fig.~\ref{fig:vc25_slices} in Appendix B for another example that shows that the effect remains striking even with halos of significantly lower mass). We clarify that we refer henceforth as "previous work" to models run with the same parameters as our full case but including neither Poisson fluctuations nor multiple scattering (above we cited previous publications related to these effects and laid out in detail their limitations). 

In the results corresponding to previous work, all simulation voxels produced a non-zero contribution to the \Lya{} intensity field, but with Poisson fluctuations taken into account, at these high redshifts only a small fraction of voxels contain star-forming halos (initially only one per voxel). The stellar \Lya{} photons produce coupling between the spin temperature and the kinetic gas temperature and produce a 21-cm absorption halo around each star-forming halo. If we increase the \Lya{} intensity (which corresponds to increasing the galaxy brightness), the 21-cm signal approaches saturation as the spin temperature approaches the kinetic temperature of the gas. Once this limit is reached near a halo, further increasing the \Lya{} intensity cannot make the nearby absorption even deeper, but it does increase the size of the coupled bubble around the halo, which makes the bubble easier to observe. Meanwhile, the disappearance of halos from many pixels (where the previous fractional numbers of halos became zero after the implementation of Poisson fluctuations) clears out the regions between the bubbles, further increasing their relative contrast.

\begin{figure*}
    \centering
 \includegraphics[width=0.7\textwidth]{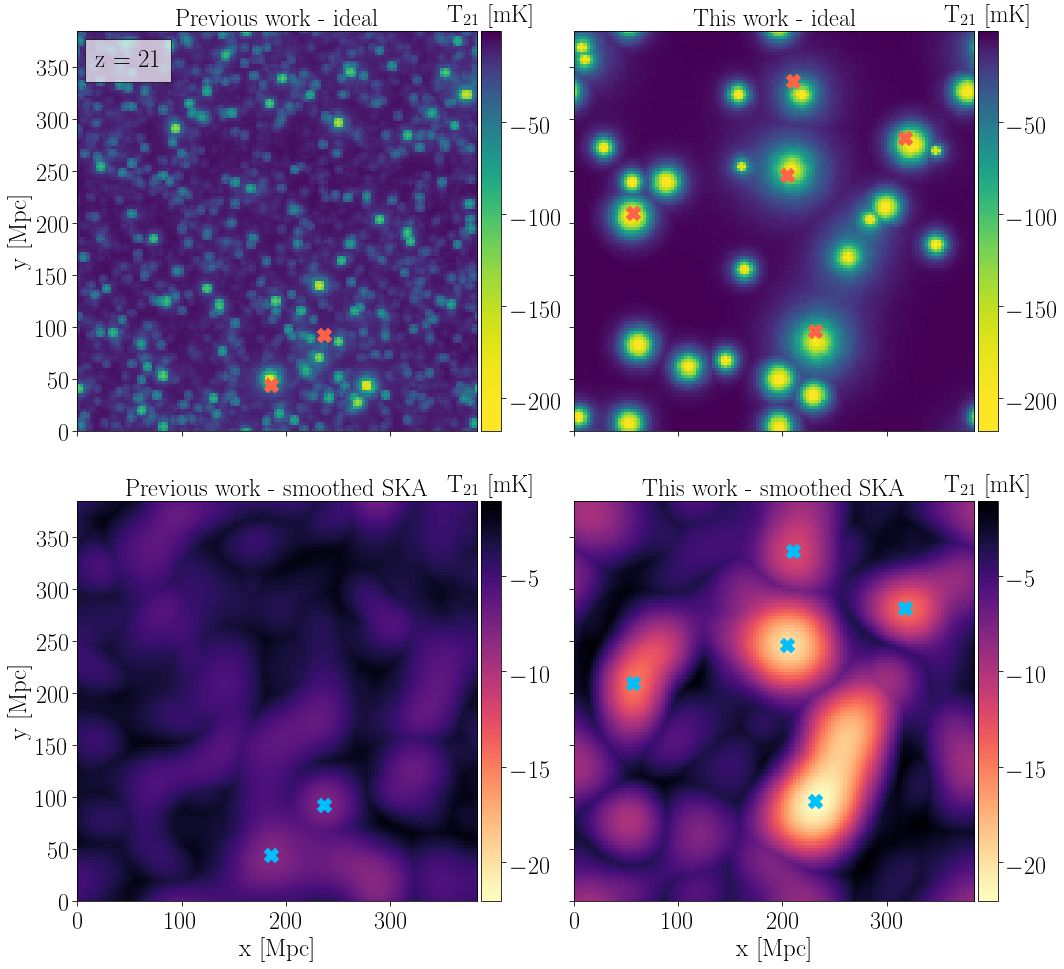}  
    \caption{{\bf Simulated images of the cosmic dawn 21-cm signal.} Since early galaxies in this model were rare, we find it useful to show a kind of projected image, defined as showing the minimum value of the signal in the direction perpendicular to the image (obtained from a simulation box that is 384 comoving Mpc on a side; each image is made of square pixels of side $3$~Mpc). All panels correspond to the same simulated volume which illustrates a model with a star-formation efficiency f$_{\star}$ = 0.1 and a minimum circular velocity V$_{\rm c}$ = 50 km/s, corresponding to a minimum star-forming halo mass of $M_{\rm min} \sim 8 \times 10^8 \; {\rm M}_{\odot}$ at the redshift shown, $z = 21$ (Fig. \ref{fig:vc25_slices} shows similarly striking effects for V$_{\rm c}$ = 25 km/s). {\bf Left panels:} Results from previous work, that is, without the effects of Poisson fluctuations and multiple scattering, shown on the scale set by the right-hand panels, for easy comparison. {\bf Right panels:} Results from this work. {\bf Top panels:} Ideal images (i.e., showing the direct simulation outputs). {\bf Bottom panels:} Projections of the same simulated volumes as in the top panel but as mock SKA images (see text); such a smoothed projection can be similarly obtained from real images. In this work, the signal is composed of large "coupled bubbles" around individual galaxies. The large size and depth of the bubbles helps them retain sufficient contrast in the mock SKA projected image to enable their detection. The locations of the $>3 \sigma$ peaks as found in the smoothed SKA box are marked in both the ideal and SKA boxes, for easy comparison. The peaks correspond to individual coupled bubbles in the ideal image, while in the SKA box there is a minor contribution smoothed in from nearby smaller bubbles. Note that some additional peaks with a lower significance can be seen in the SKA box, corresponding to smaller coupled bubbles in the ideal image. Also note that the SKA boxes are shown with respect to the cosmic mean brightness temperature, but the plotted values are negative due to our choice of showing projected minimum values.}
    \label{fig:vc50_slices}
\end{figure*}

In order to study the observational consequences of a cosmic dawn signal dominated by individual coupled bubbles as described above, we used the Square Kilometer Array \citep{koopmans15} (hereafter SKA) as an example target instrument, and created mock SKA images that account for the expected angular resolution, thermal noise, and foreground effects, all as a function of redshift. There are two common approaches to dealing with the bright foreground expected in 21-cm images. {\it Foreground removal}\/ involves modeling the foreground in order to subtract it accurately from the images (often with the help of additional observations obtained at higher resolution than needed for the cosmological 21-cm signal itself), while {\it foreground avoidance}\/ involves removing regions in $k$-space that are expected to be contaminated by the foreground. Recent work \citep{datta10, dillon14, pober14, pober15} has shown that the foreground is expected to contaminate a wedge-like region in the $k_{||}$ vs. $k_{\perp}$ plane (where we separate the wavevector to components parallel and perpendicular to the line of sight), with more foreground-free $k_{||}$ modes available at lower $k_{\perp}$ values. Since the SKA is designed to produce high-resolution deep sky images, we assume that  foreground subtraction will allow the remaining wedge of foreground avoidance to be relatively small. We refer to this reduced foreground avoidance, assumed to result from combining it with reasonably accurate foreground subtraction, as {\it mild foreground avoidance}. To the SKA images we then add a first analysis step of three-dimensional spherical smoothing, which we find to be helpful for reducing the noise \citep{quantiles} and bringing out the \Lya{} bubbles.

\begin{figure*}
    \centering
 \includegraphics[width=0.7\textwidth]{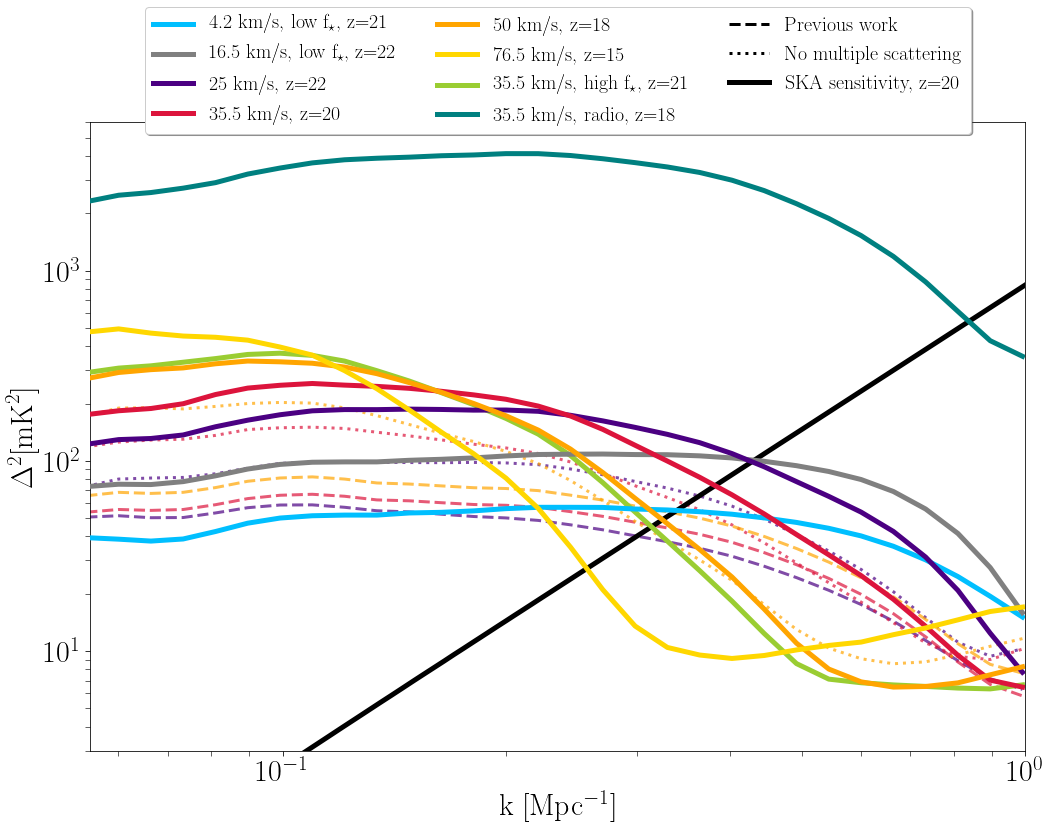}  
 
    \caption{{\bf The 21-cm power spectrum at cosmic dawn.} We show the power spectrum at the \Lya{} peak (defined as the redshift at cosmic dawn where the power spectrum at $k = 0.1 \; {\rm Mpc}^{-1}$ peaks), for various cases with or without the effects of Poisson fluctuations and multiple scattering of \Lya{} photons (solid lines and dashed lines, respectively). Unlike the gradual decline with $k$ previously expected (dashed lines, shown in three cases: $V_{\rm c} = 25$, $35.5$, and $50$~km/s), we find a strong enhancement of the fluctuations on large scales (small $k$), and a clear break in the power spectrum. The break location corresponds to the typical size of the coupled bubbles, which in turn correlates strongly with the typical brightness of individual galaxies. If the power spectrum can be measured with a noise level that approaches the expected thermal noise of the SKA (black line at $z=20$), then the break may be detectable even if star-formation occurred in halos as small as $V_{\rm c} \sim 25$~km/s (corresponding to $M_{\rm min} = 1.1 \times 10^8 \; {\rm M}_{\odot}$ at $z=20$). We also show examples of including Poisson fluctuations but not the multiple scattering of \Lya{} photons (dotted lines, corresponding to the same models as the dashed lines); both effects play a substantial role in the results shown in this work, with multiple scattering having a larger relative role in models with the lowest $M_{\rm min}$. For example, for the intermediate case of $V_{\rm c} = 35.5\ $km/s (which corresponds to ten times the minimum mass for atomic cooling, or a minimum halo mass of $M_{\rm min} \sim 3 \times 10^8 \; {\rm M}_{\odot}$ at $z=20$), the power spectrum at $k=0.1$ is enhanced by a factor of 3.8 compared to previous work, while Poisson fluctuations alone would produce an enhancement by a factor of 2.2. We also show one excess radio model motivated by the EDGES measurement (see text). For all cases shown, the power spectrum was averaged over 18 different runs of the simulation with different initial conditions and (for relevant cases) Poisson realizations.}
    \label{fig:power_spec}
\end{figure*}

A cosmic dawn signal dominated by coupled bubbles is predicted to feature prominently in the 21-cm power spectrum (Fig.~\ref{fig:power_spec}), producing a distinct power spectrum shape that is strongly correlated with the typical size of the bubbles (and thus the typical brightness of early galaxies). Coupled bubbles of size $R_{\rm bubble}$  suppress fluctuations on scales smaller than the typical bubble size, and thus result in a break in the power spectrum at $k_{\rm break} \sim 2 \pi / R_{\rm bubble}$. Meanwhile, on large scales the power spectrum is boosted compared to previous predictions by a factor that is between 2 and 7 depending on the astrophysical parameters. Thus, the signature of discrete galaxies is also a promising goal for radio arrays targeting the 21-cm power spectrum at cosmic dawn, such as the Hydrogen Epoch of Reionization Array \citep{HERA} (HERA) and the New Extension in Nan\c{c}ay Upgrading LOFAR  \citep{zarka12} (NenuFAR).

\begin{figure}
    \centering

 \includegraphics[width=0.49\textwidth]{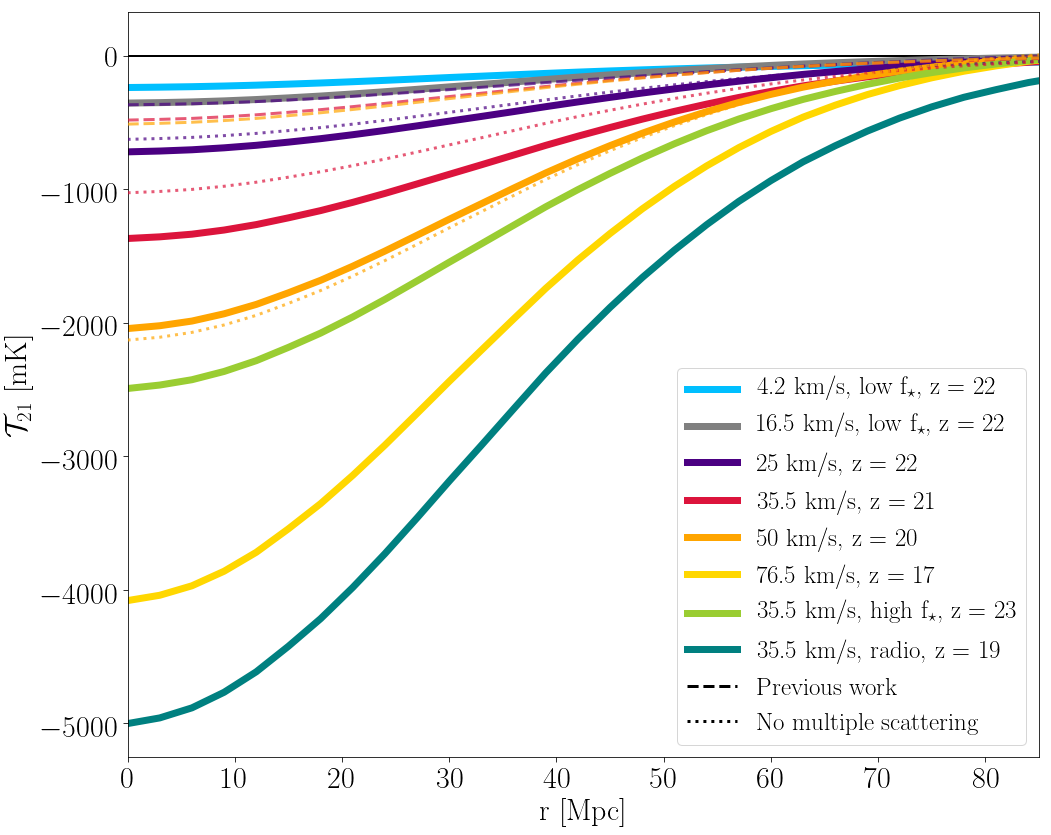}  
 
    \caption{{\bf The total peak profile at cosmic dawn.} We show the total radial profile around peaks, $\mathcal{T}_{21}(r)$, calculated from our simulated SKA data as a sum over all peaks that is then normalized per Gpc$^3$. Negative $\mathcal{T}_{21}(r)$ corresponds to the detection of coupled bubbles in 21-cm absorption on top of the SKA noise and foreground. A noise-dominated image (or any Gaussian signal) would instead give a result near zero. We show $\mathcal{T}_{21}(r)$ for the same astrophysical cases as in Fig.~\ref{fig:power_spec}. For each case, the redshift with the highest value of $\mathcal{T}_{21}(r)$ at $r=0$ is shown. We obtain significant values of $\mathcal{T}_{21}(r)$ for a wide variety of astrophysical cases. The most prominent  $\mathcal{T}_{21}(r)$ is seen for cases with higher star formation efficiencies and minimum masses for star formation. Such models produce larger and rarer (and thus higher contrast) coupled bubbles. In results corresponding to previous work, $\mathcal{T}_{21}(r)$ is weaker (in absolute value) by a factor of 2--4, with Poisson fluctuations playing the dominant role in producing the large $\mathcal{T}_{21}(r)$ obtained in this work. With Poisson fluctuations included, the prominent peaks have such a high \Lya{} intensity that their surroundings are already strongly coupled even without accounting for multiple scattering, so that the additional effect of multiple scattering on their 21-cm profile is small; however, multiple scattering consistently has a strong effect on less prominent, more typical fluctuations, as measured by the 21-cm power spectrum (Fig.~\ref{fig:power_spec}; see also Fig. \ref{fig:vc25_slices}). For each case shown, we averaged results obtained from 18 different runs of the simulation, with independent realizations of the initial conditions, Poisson fluctuations, and SKA noise. For further discussion see \cref{sec:ska_box}.}
    \label{fig:radial_profiles}
\end{figure}

While it will be intriguing to detect individual coupled bubbles in 21-cm images (as illustrated in Fig.~\ref{fig:vc50_slices}), it is important to also construct an effective statistic, to be applied to 21-cm images of cosmic dawn, that aggregates together the individual bubbles and takes advantage of this feature in order to distinguish among models. We propose the {\it total peak profile}\/ statistical probe, which measures a combination of the abundance, spatial extent, and brightness-temperature depth of the coupled 21-cm bubbles. We first detect both minima and maxima in the smoothed SKA box (see \cref{sec:ska_box}); an example of such detected peaks is shown in Fig.~\ref{fig:vc50_slices} (only peak minima are shown in the figure, for comparison with the image which shows projected minimum values of the signal). We restrict ourselves to strong peaks, defined as having a value higher (in absolute value) than $3 \sigma$, where $\sigma$ is the standard deviation of the SKA box voxel values. We calculate the radial profile around each peak that passes this threshold, and sum the profiles. The summing is done with the signed (that is, not absolute) value, in order to explicitly capture the asymmetry between maxima and minima. Indeed, any symmetric field (about its mean) would give a total result approaching zero; thus, this statistic is inherently non-Gaussian, and naturally brings out the effect of individual galaxies over thermal noise, and over any Gaussian component of the 21-cm fluctuations. Also, we sum (rather than average) these peak profiles, in order to maintain the sensitivity to the number density of peaks. In order to avoid a direct dependence on the size of the observed volume, we normalize the result by scaling it to a volume of 1~Gpc$^3$ (which corresponds almost exactly to the volume of an SKA field of view at $z=20$ with a depth of 10~MHz, or to 18 of our simulation volumes). The resulting total peak profile per volume, $\mathcal{T}_{21}(r)$, is expected to be negative during the coupling era of cosmic dawn, and measuring it as such would imply stronger minima than maxima, thus confirming the detection of coupled bubbles of 21-cm absorption above the noise level. Fig.~\ref{fig:radial_profiles} shows $\mathcal{T}_{21}(r)$ as calculated from the SKA boxes for a variety of possible parameters of the early galaxies. The expected scatter in $\mathcal{T}_{21}(r)$ due to cosmic variance and noise, for an SKA field of view, is fairly small and is shown in Appendix A (Fig.~\ref{fig:radial_profiles_scatter}).

Since 21-cm coupling requires a quite low \Lya{} intensity \citep{madau97}, it is expected to occur early enough that the observational probes considered here should be nearly unaffected \citep{cohen18} by other astrophysical radiation such as \Lya{} heating, X-ray heating or UV ionizing radiation; indeed, we have focused on signatures that occur early on, well before \Lya{} coupling approaches saturation. This means that observations at this early time depend only on the mass distribution and star-formation efficiency of halos. Higher masses of star-forming halos and higher star-formation efficiencies increase the sizes of individual \Lya{} bubbles, making it easier to detect them as well as the corresponding power spectrum break (which is moved to lower $k$). Now, higher halo masses also delay star formation and push the \Lya{} peak to a lower redshift (where observations are easier), while high efficiencies go the other way. Overall, the masses and star formation efficiencies of star-forming halos can be deduced separately given the multiple measures available, namely the amplitude and shape of the power spectrum and of the total peak profile, plus the redshifts at which these statistics peak (or, more generally, their redshift evolution). 

In both Figs.~\ref{fig:power_spec}  and \ref{fig:radial_profiles} we have included a case motivated by the recent, intriguing but not yet independently verified, EDGES measurement of the sky averaged 21-cm signal from cosmic dawn \citep{bowman18}. The EDGES measurement implies a larger than expected ratio between the background radiation and gas temperatures at cosmic dawn. This can be explained either with a lower than expected kinetic gas temperature due to a baryon - dark matter interaction \citep{barkana18,munoz18,Liu20}, or an excess radio background that raises the effective radiation temperature \citep{bowman18, feng18, fialkov19}. We include here an example of the latter model \citep{fialkov19} with parameters that are consistent with the amplitude of the absorption detected by EDGES. If such an excess radio background exists, it should give the \Lya{} bubbles a much higher contrast ($> 1000$~mK), making them even more prominent in the SKA boxes, and more easily detectable through the total peak profile $\mathcal{T}_{21}(r)$ as well as the break in the 21-cm power spectrum (which is boosted tremendously in this case). We note that here we have assumed a uniform radio background \citep{fialkov19}, but if the excess radio radiation was emitted by the same galaxies that emitted the \Lya{} radiation and created the bubbles, then the radio intensity should be higher near the galaxies thus creating an even stronger contrast for the coupled bubbles \citep{reis20c}.

\section{Summary}
\label{sec:summary}

In this work we have presented results of an upgraded simulation of the 21-cm signal from cosmic dawn, and discussed implications for planned experiments such as the SKA or the HERA. We introduced two new effects to our simulation: Poisson fluctuations in the number of galaxies, and multiple scattering of \Lya{} photons. Compared to results neglecting these effects, we found a 21-cm signal with enhanced contrast, and showing the signature of individual galaxies. In particular, the 21-cm power spectrum is enhanced by a factor of 2--7 on large scales, with a significantly different shape. Simulating SKA images we found that it should be possible to detect individual galaxies at cosmic dawn, depending on the astrophysical scenario and advancements in data analysis techniques. We also discussed the {\it total peak profile} - an effective statistic that could be applied to future observations to distinguish between models.

For simplicity, in this work we have assumed a constant value of the star formation efficiency f$_{\star}$ for all star-forming halos. While this approach is common, in reality we expect significant scatter in the f$_{\star}$ value among galaxies, due to different merger and accretion histories, and as found in simulations \citep{xu16}. We have tested the effect of a galaxy to galaxy variance in the star formation efficiency and found that a significant variance can strongly enhance the coupled bubble signature in the cosmic dawn signal. Even for a lower {\it average}\/ star formation efficiency than we have assumed, a variance in this parameter should still result in a few galaxies bright enough to produce large coupled bubbles that can be detected by the SKA. This highlights again the fact that the \Lya{} bubble cosmic dawn signal predicted here is produced by individual galaxies and affected by small number statistics at the tail of the brightness distribution. This is in contrast to previous work predicting a signal dominated by large scale structure and determined by the average properties of the galaxy population. Our results are a boon to planned 21-cm observations of cosmic dawn, as they predict favorable observational targets in the form of an enhanced 21-cm power spectrum and a strongly non-Gaussian 21-cm signal, even if most star-forming halos were small as is generally expected. Finally, we note that the novel effects investigated here also affect the global 21-cm signal (due to the non-linearity of the 21-cm fluctuations) but only marginally, at the few to ten percent level at the redshifts investigated here.

\section*{Acknowledgments}

We acknowledge the usage of the DiRAC HPC. AF was supported by the Royal Society University Research Fellowship. This project was made possible for I.R.\ and R.B.\ through the support of the ISF-NSFC joint research program (grant No.\ 2580/17). 

This research made use of:
 {\fontfamily{cmtt}\selectfont  SciPy}  \citep[including {\fontfamily{cmtt}\selectfont  pandas} and {\fontfamily{cmtt}\selectfont NumPy, }][]{2020SciPy, numpy},  {\fontfamily{cmtt}\selectfont IPython}  \citep[][]{perez07}, {\fontfamily{cmtt}\selectfont matplotlib}  \citep[][]{hunter07}, {\fontfamily{cmtt}\selectfont Astropy}  \citep[][]{astropy-collaboration13}, {\fontfamily{cmtt}\selectfont Numba}  \citep{lam15},  the SIMBAD database  \citep[][]{wenger00}, and the NASA Astrophysics Data System Bibliographic Services.

\section*{Data availability}
No new data were generated or analysed in support of this research.

\bibliographystyle{mnras}
\bibliography{dgcd}

\appendix

\section{SKA boxes and peak detection.} 
\label{sec:ska_box}
In order to simulate mock SKA images, we apply the following procedure
to the ideal images that are directly output from our semi-numerical simulations: (i) We smooth the ideal images with a two-dimensional Gaussian that corresponds approximately to the effect of SKA resolution. (ii) We add to the image a random realization of the SKA thermal noise, a pure Gaussian noise smoothed with the SKA resolution. The amplitude of the noise depends on the redshift and the SKA resolution. (iii) To account for the effects of the foreground, we adopt mild foreground avoidance. 

The SKA is designed to be used at a range of different resolutions, depending on which range of baselines are included in the image. Higher resolution is not necessarily better as it comes with higher noise. Here we use a radius (corresponding approximately to half the full width at half maximum of the point spread function) of $R_{\rm SKA} = 20$ Mpc, since 10~Mpc seems too noisy while 40~Mpc appears to wipe out too much of the signal \citep{quantiles}; it may be useful to explore more systematically how the results of this work vary with this additional parameter, the resolution.

The strength of the SKA thermal noise (for a frequency depth corresponding to 3~comoving Mpc, and assuming a 1000~hour integration by the SKA) is approximately given by \citep{koopmans15,quantiles}
\begin{equation}
\sigma_{\rm thermal} = a\, \left(\frac{1+z}{17}\right)^b,
\end{equation}
where the parameters $a$ and $b$ depend on the resolution used. For $R_{\rm SKA} = 20$ Mpc, which we use here, $a = 4$~mK and $b = 5.1$ gives a good fit for $z > 16$ (mainly of interest in this work), while for lower redshifts, $b = 2.7$.

As noted in the main text, foreground avoidance is the approach in which regions in $k$-space that are expected to be contaminated by foreground are removed. Assuming that the SKA will enable
a first step of reasonably accurate foreground subtraction, our mild foreground avoidance assumes that the remaining wedge-like region will be relatively small. We assume that the remaining wedge that is still contaminated by foreground is given by $k_{||} < C(z) k_{\perp}$ , where \citep{dillon14, jensen15}
\begin{equation}
    C(z) =  \frac{D_{\rm M}(z) \sqrt{\Omega_{m} (1+z)^3 + \Omega_{\Lambda}}}{(1+z) \times  c/H_{0}}\ \sin{R_{\rm FoV}}\ .
\end{equation}
Here $R_{\rm FoV}$ is the angular radius of the field of view (for which we use the redshift-dependent value expected for the SKA) and $D_{\rm M}$ is the (transverse) comoving distance. The model used here corresponds to an "optimistic model" from previous work \citep{pober14}. The real shape of the wedge is still under debate, and depends on advancements in analysis techniques. It is possible that foreground modeling and subtraction techniques will prove to produce better results than the foreground avoidance technique applied in this work. 
 
 To remove modes inside the wedge, we Fourier transform the 21-cm image (after two-dimensional smoothing and the addition of thermal noise), remove all modes inside the wedge, and inverse Fourier transform. To obtain the final SKA box which we use in the analysis, we further apply three-dimensional smoothing with a spherical top-hat of radius 20~Mpc (same as the radius of the SKA angular resolution), as a first step in the analysis (not part of producing a mock SKA image). Since we are interested in spherically-averaged peak profiles, it is sensible to add this smoothing in order to even out the differences between the angular directions (which suffer the instrumental resolution smoothing) and the line of sight. More importantly, since the \Lya{} bubbles are fairly large, we find that this step smooths out the thermal noise more than the signal, and thus brings out the bubbles \citep{quantiles}. 
 
 Given the smoothed mock SKA boxes, we detect all maxima and minima, in order to calculate the total radial peak profile $\mathcal{T}_{21}(r)$. We then use a standard local peak detection algorithm implemented in \texttt{scikit-image}\footnote{scikit-image.org}. This simple approach for detecting the peaks is applicable here, since the SKA images are smooth (especially after the additional three-dimensional smoothing, which makes things more robust). We validated the results by visual inspection. 
 
\begin{figure}
    \centering
 
 \includegraphics[width=0.49\textwidth]{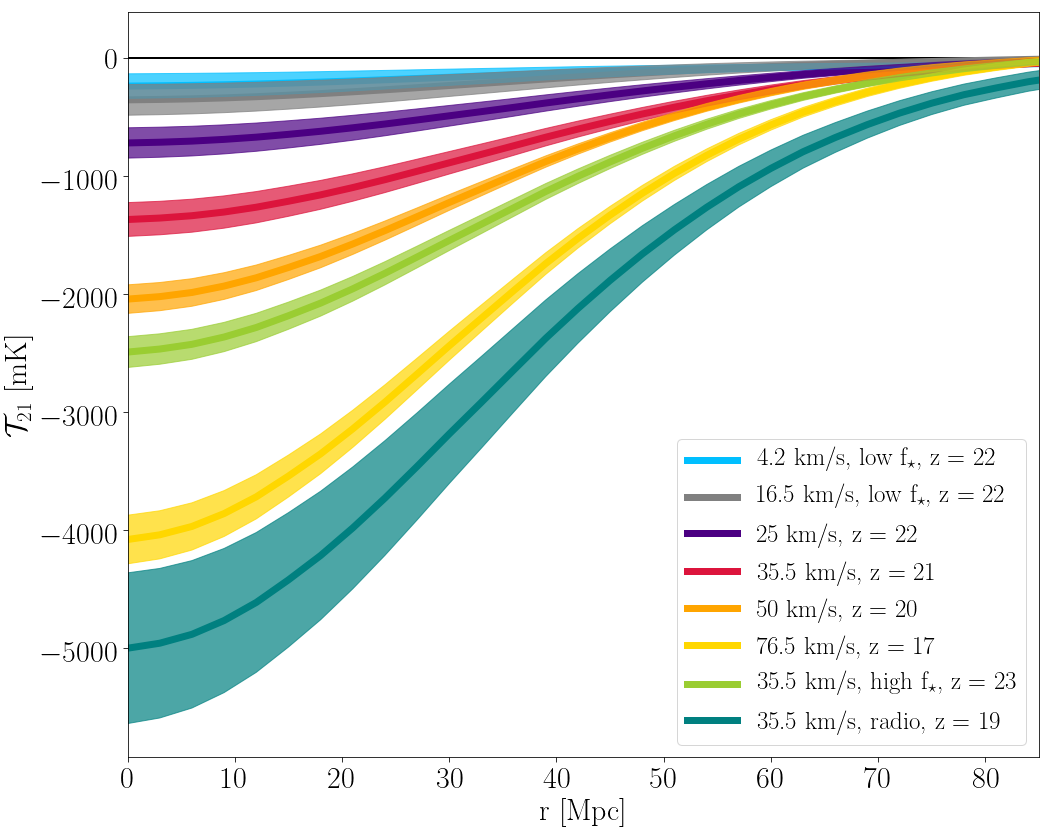}  
 
    \caption{{\bf Scatter in the total peak profile.} This plot is the same as Fig.~\ref{fig:radial_profiles}, but it shows also the scatter (only for this work, with all the new physical effects included). The scatter originates from both cosmic variance (which in turn originates from both the density field realization, and the Poisson realization of the dark matter halos), and SKA noise. For each case shown, we calculated the standard deviation of the radial profile using our 18 simulation boxes, and then scaled the result to correspond to the scatter in an SKA field of view. This is shown for each case with a colored region centered around the mean value.}
    \label{fig:radial_profiles_scatter}
\end{figure}

As shown in the main text, $\mathcal{T}_{21}(r)$ is a useful probe of the 21-cm signal. We note here a few additional points regarding this quantity. Its scatter as expected for an SKA field of view is shown in Fig. \ref{fig:radial_profiles_scatter}; the scatter is small enough that it should not make it hard to distinguish among models with significantly different values of the minimum circular velocity $V_c$ of star-forming halos, except for the very lowest values (16.5~km/s and below). Also, $\mathcal{T}_{21}(r)$ is a relative quantity, in the sense that it is sensitive to the number of peaks higher than three standard deviations of the signal. The standard deviation is sometimes dominated by thermal noise, but in many cases the (smoothed) original signal and foreground  effects play more important roles. For this reason we have used the standard deviation of the total mock signal, and not just that of the thermal noise, in the definition of $\mathcal{T}_{21}(r)$. Deeper peaks produce a higher standard deviation through the foreground residuals resulting from mild foreground avoidance. In some cases the positive peaks in the SKA boxes are not due to thermal noise, but due to these artifacts originating from foreground avoidance, with an amplitude that tends to correlate with the strength of the 21-cm signal. All of this reduces the differences among models with different parameters. It is possible that a more elaborate foreground-removal algorithm would improve this behavior; as one example idea, the highest detected peak can be fitted, and then the foreground residuals resulting from it can be removed approximately from the image before the analysis continues onward, peak by peak. Finally, we note that while we have found that a threshold of three standard deviations works well, this is really an additional parameter of this statistic that can be further studied.

\section{Low $V_c$ example}
\begin{figure*}
    \centering
 \includegraphics[width=0.95\textwidth]{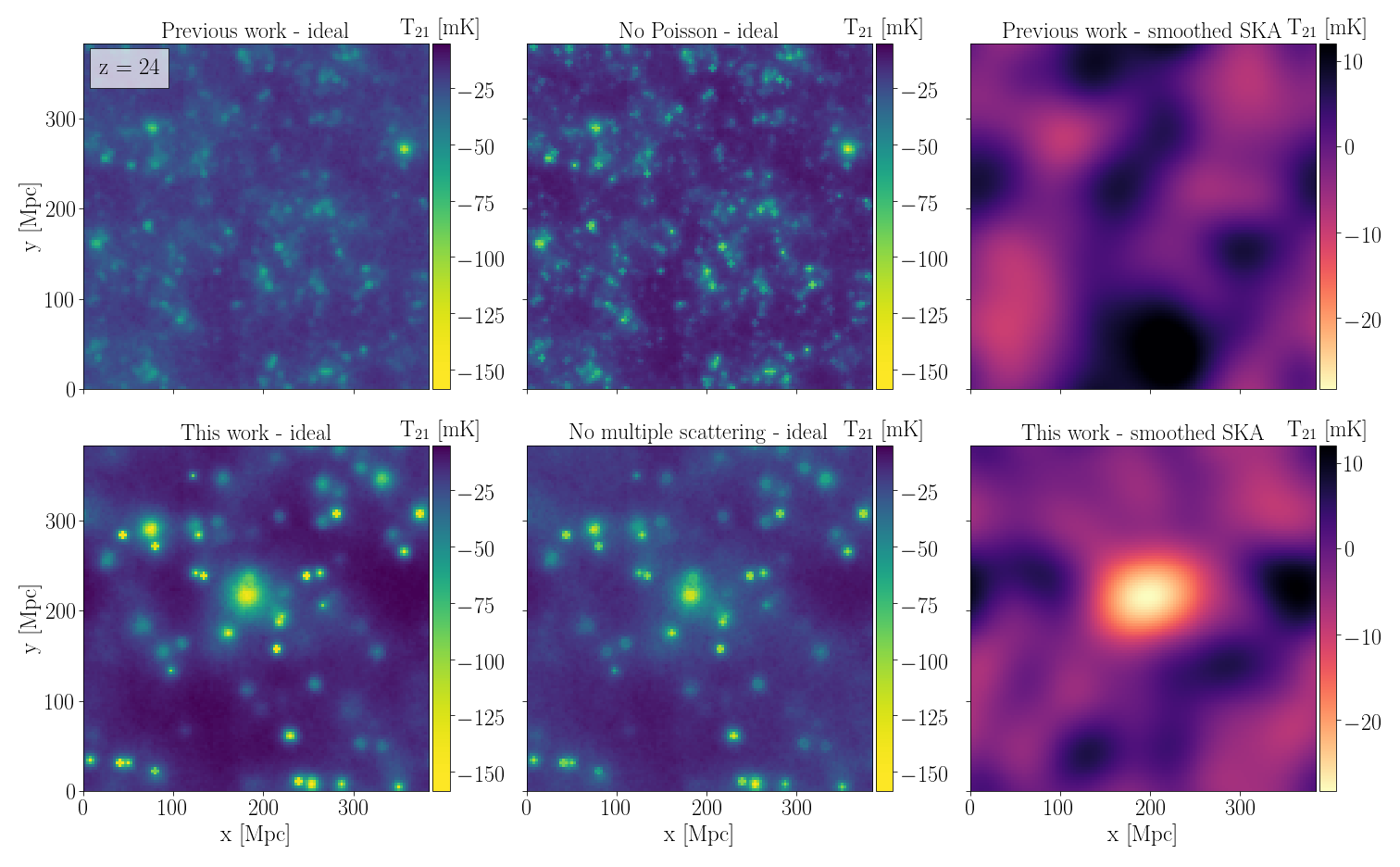}  
    \caption{{\bf Simulated 21-cm images of cosmic dawn for a model in which small galaxies dominate.} This is similar to Fig.~\ref{fig:vc50_slices} but for V$_{\rm c}$ = 25 km/s (corresponding to a minimum star-forming halo mass of $M_{\rm min} \sim 8 \times 10^7 \; {\rm M}_{\odot}$ at $z = 24$) and f$_{\star}$ = 0.1. Given the higher galaxy density in this model, here we show a thin (one-voxel) slice and not a projection as in Fig.~\ref{fig:vc50_slices}. For physical understanding, we also show intermediate cases (for the ideal image) that include only multiple scattering (No Poisson) or only Poisson fluctuations (No multiple scattering). Multiple scattering concentrates the \Lya{} photons near sources, making the 21-cm peaks brighter (in absolute value) and the low-intensity regions darker. Even in an astrophysical scenario with many low-mass star-forming galaxies, Poisson fluctuations in the numbers of rare high-mass galaxies can still produce a large, observable, effect. In this example a coupled bubble produced (mainly) by an individual massive halo ($\sim 9\times 10^8 {\rm M}_{\odot}$) is seen. The coupled bubble is visible on top of a 21-cm signal otherwise dominated by low-mass galaxies, and is prominently seen in the SKA image (while nothing similar is seen in the previous-work version). For the astrophysical model shown here, a few such \Lya{} bubbles are expected within an SKA field of view.}
    \label{fig:vc25_slices}
\end{figure*}

In Fig. \ref{fig:vc25_slices} we show simulated images of cosmic dawn, similarly to Fig. \ref{fig:vc50_slices}, only for a model with lower mass galaxies. This example shows that the signature of individual massive galaxies can possibly be seen even in models with many star forming low mass galaxies.

% Don't change these lines
\bsp	% typesetting comment
\label{lastpage}
\end{document}